\documentclass[twocolumn,prb,showpacs,floatfix]{revtex4}

\usepackage{graphicx}
\usepackage{dcolumn}
\usepackage{bm}
\usepackage{amsmath}

\begin{document}

\title{Quantum dots in high magnetic fields:
Rotating-Wigner-molecule versus composite-fermion approach}

\author{Constantine Yannouleas}
\email{Constantine.Yannouleas@physics.gatech.edu}
\author{Uzi Landman}
\email{Uzi.Landman@physics.gatech.edu}

\affiliation{School of Physics, Georgia Institute of Technology,
             Atlanta, Georgia 30332-0430}

\date{April 2003} 

\begin{abstract}
Exact diagonalization results are reported for the lowest rotational 
band of $N=6$ electrons in strong magnetic fields in the range
of high angular momenta $ 70 \leq L \leq 140 $ (covering the corresponding 
range of fractional filling factors $1/5 \geq \nu \geq 1/9$). A detailed 
comparison of energetic, spectral, and transport properties (specifically, 
magic angular momenta, radial electron densities, occupation 
number distributions, overlaps and total energies, and exponents of 
current-voltage power law) shows that the recently discovered 
rotating-electron-molecule wave functions [Phys. Rev. B {\bf 66}, 115315 
(2002)] provide a superior description compared to the 
composite-fermion/Jastrow-Laughlin ones.
\end{abstract}


\maketitle

\section{Introduction}

Two-dimensional (2D) $N$-electron systems (with a small finite $N$) in strong 
magnetic fields ($B$) have been the focus of extensive 
theoretical investigations in the last twenty 
years.\cite{lau1,lau2,gir,jai3,mac,jai1,jai4,mk,sek,%
mak1,rua,yl2,mak2,cre2m,yl5,ron1m,sba}
The principal motivations for these research activities are: 
(I) The early realization \cite{lau1,lau2} 
that certain special states of few electron systems are relevant\cite{haw} 
through appropriate analogies to the physics
of the fractional quantum Hall effect (FQHE), observed in the infinite 2D 
electron gas; (II) The unavoidable necessity, due to computer limitations, to
test proposed model wave functions for the FQHE through numerical calculations
for finite-size systems; (III) The recent progress in nanofabrication 
techniques at semiconductor interfaces that has allowed experiments on 2D 
quantum dots (QD's), with refined control of their size, shape, and number of 
electrons\cite{ash,tar,kou} (down to a few electrons).

The physics of such systems (i.e., QD's in high $B$), is most often described
with the use of composite-fermion\cite{jai3}/Jastrow-Laughlin\cite{lau1}  
(CF/JL) analytic trial wave functions in the complex plane. However, it is 
well known that the thematic framework of the CF/JL approach is built on the 
so-called Jastrow correlations associated with a particular short-range 
interparticle repulsion.\cite{hk} In a recent paper,\cite{yl5} using as a 
thematic basis the picture of collectively rotating electron (or Wigner) 
molecules (REM's), we have derived a different class of analytic and 
parameter-free trial wave functions. The promising property of these REM wave 
functions is that, unlike the CF/JL ones, they capture the all-important 
correlations arising from the long-range character of the Coulomb force. 

In this paper, we present an in-depth assessment of the CF/JL and REM trial 
wave functions regarding their ability to approximate the exact wave functions
in the case of QD's (this case is often referred to as the ``disk geometry''
in the FQHE literature). First systematic exact diagonalization (EXD) results 
are reported here for the lowest rotational band of $N=6$ electrons in strong 
magnetic fields in the range of high angular momenta $ 70 \leq L \leq 140 $ 
(covering the corresponding range of fractional filling factors\cite{note4} 
$1/5 \geq \nu \geq 1/9$). A detailed comparison (addressing five properties, 
i.e., prediction of magic angular momenta, radial electron densities, 
occupation number distributions, overlaps and total energies, and exponents of
current-voltage power law) shows that the REM wave functions yield a superior
description to that obtained through the composite-fermion/Jastrow-Laughlin 
ones.

The plan of this paper is as follows: Section II presents an outline of the
REM theory, while section III focuses on a brief review of the 
composite-fermion approach. Exact-diagonalization results and comparisons
with the CF/JL and REM wave functions are presented in section IV.
Finally, our results are summarized in section V.

\section{Outline of REM theory}

In the last eight years, and in particular since 1999 [when it was
demonstrated\cite{yl1} that Wigner crystallization is related to 
symmetry breaking at the {\it unrestricted\/} Hartree-Fock (UHF) mean-field 
level], the number of publications$^{8-17, 24-40}$ 
addressing the formation and properties of Wigner (or electron) molecules in 
2D QD's and quantum dot molecules has grown steadily. A consensus has 
been reached that rotating electron molecules are formed both in zero 
\cite{yl1,cre1,gra1,yl2,gra2,loz1,man1,gra3,pee1,loz2,yl3,yl4,loz3,%
mik1,mik2,mik3,nie1,yl8} and high\cite{mk,sek,mak1,rua,yl2,mak2,cre2m,yl5,%
ron1m,sba} magnetic fields.

At $B=0$, formation of REM's in QD's is analogous to Wigner crystallization 
in infinite 2D media, i.e., when the strength 
of the interelectron repulsion relative 
to the zero-point kinetic energy ($R_W$) exceeds a certain critical value, 
electrons spontaneously crystallize around sites forming geometric
molecular structures. At high magnetic fields, the formation of Wigner 
molecules may be thought of as involving a two-step crystallization process: 
(I) the localization of electrons results from the shrinkage of the orbitals
due to the increasing strength of the magnetic field; (II) then, 
even a weak interelectron Coulomb repulsion is able to arrange the localized
electrons according to geometric molecular structures (thus this process is 
independent of the value of $R_W$). It has been found\cite{mk,mak1,yl2} 
that the molecular structures at high $B$ coincide with the equilibrium 
configurations at $B=0$ of $N$ classical point charges.\cite{bed,bol} 

Due to the finite number, $N$, of electrons, however,
there are two crucial differences between the REM and the bulk Wigner crystal.
Namely, (I) the crystalline structure is that of the equilibrium 2D 
configuration of $N$ classical point charges, and thus consists of nested 
polygonal rings; \cite{note24} 
(II) the Wigner molecules rotate as a whole (collective 
rotations) in analogy with the case of 3D natural molecules.

A most striking observation concerning the REM's is that their formation
and properties have been established with the help of traditional {\it ab 
initio\/} many-body methods, i.e., exact diagonalization,
\cite{cre1,yl3,mik3,sek,mak1,rua,ron1m}  
quantum Monte Carlo\cite{gra1,loz1,loz2,nie1} (QMC), and the systematic 
controlled hierarchy\cite{yl1,yl2,yl4,loz3,mk,yl5,sba,yl8} of approximations 
involving the UHF and subsequent post-Hartree-Fock
methods. This contrasts with the case of the CF/JL wave functions, 
which were inspired through ``intuition-based guesswork''.

In spite of its firm foundation in many-body theory, however, the REM picture
has not, until recently, successfully competed with the CF/JL picture; indeed 
many research papers\cite{jai2,gr,yang,tau,man2,rez,rez2,nie2} and 
books\cite{haw} describe the physics of QD's 
in high magnetic fields following exclusively
notions based on CF/JL functions, as expounded in 1983 (see Ref.\ 
\onlinecite{lau1}) and developed in detail in 1995 in Ref.\ 
\onlinecite{jai1} and Ref.\ \onlinecite{jai4}. 
We believe that one of the main obstacles for more 
frequent use of the REM picture has been the lack of analytic correlated wave 
functions associated with this picture. This situation, however, has 
changed with the recent explicit derivation of such REM wave functions.
\cite{yl5}

The approach used in Ref.\ \onlinecite{yl5} for constructing the REM functions
in high $B$ consists of two-steps: First the
breaking of the rotational symmetry at the level of the single-determinantal
unrestricted Hartree-Fock approximation yields states
representing electron molecules (or finite  crystallites, also referred to
as Wigner molecules, see Ref.\ \onlinecite{yl1} and Ref.\ \onlinecite{yl2}). 
Subsequently the rotation of the electron molecule is 
described through restoration of the circular symmetry via post Hartree-Fock 
methods, and in particular Projection Techniques.\cite{rs} Naturally, 
the restoration of symmetry goes beyond the single determinantal mean-field 
description and yields
multi-determinantal wave functions. For QD's, we have shown that the method
of symmetry restoration is applicable to both the zero\cite{yl4,yl8} and 
high\cite{yl5} magnetic-field cases. 

In the zero and low-field cases, the broken symmetry UHF orbitals need to be 
determined numerically, and, in addition, the restoration of the total-spin
symmetry needs to be considered for unpolarized and partially polarized cases.
The formalism and mathematical details of this procedure at $B=0$ have been 
elaborated in Ref.\ \onlinecite{yl4} (see also Ref.\ \onlinecite{yl6} and
Ref.\ \onlinecite{yl7}) for the restoration of the total spin in the case of 
quantum dot molecules).

In the case of high magnetic fields, one can specifically consider the limit
when the confining potential can be neglected compared to the confinement
induced by the magnetic field. Then, assuming a symmetric gauge, the UHF 
orbitals can be represented \cite{yl5,mz} by displaced Gaussian analytic 
functions, centered at different positions $Z_j \equiv X_j+ \imath Y_j$ 
according to the equilibrium configuration of $N$ classical point 
charges\cite{bed,bol} arranged at the vertices of nested regular polygons 
(each Gaussian representing a localized electron). Such displaced Gaussians 
are written as (here and in the following $\imath \equiv \sqrt{-1}$)
\begin{eqnarray}
u(z &,& Z_j) = (1/\sqrt{\pi}) \nonumber \\
&\times& \exp[-|z-Z_j|^2/2] \exp[-\imath (xY_j-yX_j)],
\label{gaus}
\end{eqnarray}
where the phase factor is due to the gauge invariance. $z \equiv x+\imath y$
(see Ref.\ \onlinecite{note5}), and all lengths are in dimensionless units of 
${l_B}\sqrt{2}$ with the magnetic length being $l_B=\sqrt{\hbar c/eB}$. 

In Ref.\ \onlinecite{yl5}, we used these analytic orbitals to first construct 
the broken symmetry UHF determinant, $\Psi^{\text{UHF}}_N$, and then proceeded
to derive analytic expressions for the many-body REM wave functions by 
applying onto $\Psi^{\text{UHF}}_N$ an appropriate projection 
operator\cite{yl5} ${\cal O}_L$ that restores the circular symmetry and 
generates {\it correlated\/}\cite{note23} wave functions
with good total angular momentum $L$. These REM wave functions can be easily 
written down\cite{yl5} in second-quantized form for any classical polygonal 
ring arrangement $(n_1,n_2,...)$ by following certain simple rules for 
determining the coefficients of the determinants $D(l_1,l_2,...,l_N) \equiv 
{\text{det}}[z_1^{l_1},z_2^{l_2}, \cdot \cdot \cdot, z_N^{l_N}]$,
where the $l_j$'s denote the angular momenta of the individual electrons.
Since we will focus here on the case of $N=6$ and $N=3$ electrons, we list for
completeness the REM functions associated with the $(0,N)$ and $(1,N-1)$ ring 
arrangements, respectively [here $(0,N)$ denotes a regular polygon with $N$
vertices, such as an equilateral triangle or a regular hexagon, and $(1,N-1)$
is a regular polygon with $N-1$ vertices and one occupied site in its center], 
\begin{eqnarray}
\Phi_L (0,N) && =
 \sum^{l_1 + \cdot \cdot \cdot +l_N=L}%
_{0 \leq l_1<l_2< \cdot \cdot \cdot <l_N}
\left( \prod_{i=1}^N l_i! \right)^{-1}  \nonumber \\
&& \times \left( \prod_{1 \leq i < j \leq N} 
\sin \left[\frac{\pi}{N}(l_i-l_j)\right] \right)  \nonumber \\
&& \times \; D(l_1,l_2,...,l_N)
\exp(-\sum_{i=1}^N z_i z_i^*/2),
\label{phi1}
\end{eqnarray} 
with 
\begin{equation}
L=L_0+Nm, \;\; m=0,1,2,3,...,
\label{lwm0n}
\end{equation}
and
\begin{eqnarray}
\Phi_L (1,&N&-1) = 
\sum^{l_2+ \cdot \cdot \cdot +l_N=L}%
_{1 \leq l_2 < l_3 < \cdot\cdot\cdot < l_N}
\left( \prod_{i=2}^N l_i! \right)^{-1} \nonumber \\
&& \times \left( \prod_{2 \leq i < j \leq N} 
\sin \left[\frac{\pi}{N-1}(l_i-l_j)\right] \right)  \nonumber \\
 \nonumber \\
&& \times \; D(0,l_2,...,l_N)
\exp(-\sum_{i=1}^N z_i z_i^*/2),
\label{phi2}
\end{eqnarray}
with 
\begin{equation}
L=L_0+(N-1)m, \;\; m=0,1,2,3,...,
\label{lwm1nm1}
\end{equation}
where $L_0=N(N-1)/2$ is the minimum allowed total angular momentum for
$N$ (polarized) electrons in high magnetic fields.

Notice that the REM wave functions [Eq.\ (\ref{phi1}) and Eq,\ (\ref{phi2})]
vanish identically for values of the total angular momenta outside the 
specific values given by Eq.\ (\ref{lwm0n}) and Eq.\ (\ref{lwm1nm1}), 
respectively.

\section{Outline of Composite-Fermion theory}

According to the CF picture,\cite{jai1} the many body wave functions in high
magnetic fields that describe $N$-electrons in the disc geometry (case of 2D 
QD's) are given by the expression,
\begin{equation}
\Phi^{\text{CF}}_L (N) = 
{\cal P_{\text{LLL}}} \prod_{1 \leq i < j \leq N} (z_i - z_j)^{2m} 
\Psi^{\text{IPM}}_{L^*},   
\label{cfeqs}
\end{equation}
where $z=x+\imath y$ and $\Psi^{\text{IPM}}_{L^*}$ is the Slater determinant 
of $N$ {\it non-interacting\/} electrons of total angular momentum $L^*$; it
is constructed according to the Independent Particle Model (IPM) from the
Darwin-Fock\cite{df} orbitals $\psi_{p,l}(z)$, where $p$ and $l$ are the
number of nodes and the angular momentum, respectively [for the values of $p$ 
and $l$ in the $n$th Landau level in high $B$, see the paragraph following
equation (\ref{ltol*}) below].

The Jastrow 
factor in front of $\Psi^{\text{IPM}}_{L^*}$ is introduced to represent the 
effect of the interelectron Coulombic interaction. In the CF literature, this 
assumption is often described by saying that ``the Jastrow factor binds $2m$ 
vortices to each electron of $\Psi^{\text{IPM}}_{L^*}$ to convert it into a 
composite fermion''. 

The single-particle electronic orbitals in the Slater determinant 
$\Psi^{\text{IPM}}_{L^*}$ are not restricted to the lowest Landau level (LLL).
As a result, it is necessary to apply a projection operator 
${\cal P_{\text{LLL}}}$ to guarantee that the CF wave function lies in the
LLL, as appropriate for $B \rightarrow \infty$.

Since the CF wave function is an homogeneous polynomial in the electronic
positions $z_j$'s, its angular momentum $L$ is related to the non-interacting
total angular momentum $L^*$ as follows,
\begin{equation}
L = L^* + m N(N-1) = L^* +2mL_0.
\label{ltol*}
\end{equation}

There is no reason to {\it a priori\/} restrict the Slater determinants
$\Psi^{\text{IPM}}_{L^*}$ to a certain form, but according to Ref.\ 
\onlinecite{jai1}, such a restriction is absolutely necessary in order to
derive systematic results. Thus following Ref.\ \onlinecite{jai1}, henceforth,
we will restrict the non-interacting $L^*$ to the range $-L_0 \leq L^* \leq
L_0$, and we will assume that the Slater determinants $\Psi^{\text{IPM}}_{L^*}$
are the so-called compact ones. Let $N_n$ denote the number of electrons in the
$n$th Landau Level (LL) with $\sum_{n=0}^t N_n = N$; $t$ is the index
of the highest occupied LL and all the lower LL's with $n \leq t$ are assumed 
to be occupied.  The compact determinants are defined as those in 
which the $N_n$ electrons occupy contiguously the single-particle orbitals 
(of each $n$th LL) with the lowest angular momenta, 
$l=-n, -n+1, ..., -n+N_n-1$ $[p+(|l|-l)/2 = n]$. The compact Slater 
determinants are usually denoted as $[N_0, N_1, ..., N_t]$, and the 
corresponding total angular momenta are given by 
$L^* =(1/2) \sum_{s=0}^t N_s(N_s-2s-1)$.

Most important for our present study is the fact that the Jastrow-Laughlin
wave functions with angular momentum $L=(2m+1)L_0$ [corresponding
to fractional filling factors $\nu =L_0/L=1/(2m+1)$],
\begin{equation}
\Phi^{\text{JL}}_L (N) = \prod_{1 \leq i < j \leq N} (z_i - z_j)^{2m+1} 
\exp \left( - \sum_{k=1}^N z_k z_k^*/2 \right),
\label{jlwfs}
\end{equation}
are a special case of the CF functions for $L^*=L_0$, i.e.,
\begin{equation}
\Phi^{\text{JL}}_{L}(N) = \Phi^{\text{CF}}_L (N; L^*=L_0),\;\; L=(2m+1)L_0.
\label{jlcf}
\end{equation}
Note that for $L^*=L_0$, all the non-interacting electrons occupy contiguous
states in the LLL ($n=0$) with $l=0,1,...,N-1$.

The CF/JL wave functions [equations (\ref{cfeqs}) and (\ref{jlwfs})] are 
represented by compact, one-line mathematical expressions, which however are 
not the most convenient for carrying out numerical calculations. Numerical 
studies of the CF/JL functions usually employ sophisticated Monte Carlo 
computational techniques. The REM wave functions, on the other hand, are by 
construction expressed in second-quantized (superposition of Slater 
determinants) form, precisely like the wave functions from exact 
diagonalization, a fact that greatly simplifies the numerical work. In the 
numerical calculations involving JL wave functions in this paper, we have 
circumvented the need to use Monte Carlo techniques, since we were able to 
determine the Slater decomposition\cite{note1} of the JL states with the help 
of the symbolic language MATHEMATICA.\cite{math} 

We stress again that, unlike the REM functions, the CF/JL wave functions
have not been derived microscopically, i.e., from the many-body Schr\"{o}dinger
equation with interelectron Coulombic repulsions. Attempts have been made to
justify them {\it a posteriori\/} by pointing out that their overlaps with 
exact wave functions are close to unity or that their energies are close to
the exact energies. However, we will show below that this agreement is limited
to rather narrow ranges of filling factors between $1 \geq \nu \geq 1/3$ or to
small electron numbers $N$; as soon as one extends the comparisons to a 
broader range of $\nu$'s for $N \geq 6$, as well as to other quantities like 
electron densitiess and occupation number distributions, this agreement 
markedly deteriorates.

\section{Exact diagonalization results and comparisons}

In the case of high magnetic fields, the Hilbert space for 
exact-diagonalization calculations can be restricted to the LLL and many such 
calculations have been reported\cite{sek,mak1,jai2,haw,yang,rez,lau2,gir,sto,%
mac,kash,tsi,kasn} in the past twenty years. However, for $N \geq 5$, such EXD 
studies have been restricted to angular momenta corresponding to the rather 
narrow range of fillings factors $1 \geq \nu \geq 1/3$. 

In this paper, we have performed systematic EXD calculations in the LLL for 
$N=6$ electrons covering the much broader range of fillings factors 
$1 \geq \nu \geq 1/9$; such a range corresponds to angular momenta 
$15 \leq L \leq 140$ (note that for $\nu=1/3$ one has $L=45$). Of crucial 
importance for extending the calculations to such large $L$'s has been our 
use of Tsiper's\cite{tsi2} analytic 
formula for calculating the two-body matrix elements of the Coulomb 
interelectron repulsion; this formula expresses the matrix elements as finite 
sums of positive terms. Earlier analytic formulas\cite{gir} suffered from
large cancellation errors due to summations over alternating positive and
negative terms. At the same time, Tsiper's formula is computationally faster
compared to the slowly-convergent series of Ref.\ \onlinecite{sto}.

For the solution of the large scale, but sparse, Coulomb eigenvalue problem, 
we have used the ARPACK computer code.\cite{arp} For a given $L$, the Hilbert 
space is built out of Slater determinants,
\begin{equation}
D(l_1,l_2,...,l_N) \exp(-\sum_{i=1}^N z_i z_i^*/2),  
\label{sld}
\end{equation}
with
\begin{equation}
l_1 < l_2 < ... < l_N,\;\;\;\;\; \sum_{k=1}^N l_k = L,
\label{sldc}
\end{equation}
and its dimensions are controlled by the maximum allowed single-particle
angular momentum $l_{\text{max}}$, such that $l_k \leq l_{\text{max}}$,
$1 \leq k \leq N$. We have used $l_{\text{max}}=
l^{\text{JL}}_{\text{max}}+5 = 10(m+1)$ (see Ref.\ \onlinecite{note1}
for the definition of $l^{\text{JL}}_{\text{max}}$) for each group of angular 
momenta $L$ corresponding to the range $1/(2m+1) \leq \nu < 1/(2m-1)$, 
$m=1,2,3,4$. For example, for $L=105$, $l_{\text{max}}=40$ and 
the dimension of the Hilbert space is 56115; for $L=135$, $l_{\text{max}}=50$
and the size of the Hilbert space is 187597. By varying $l_{\text{max}}$, we 
have checked that this choice produces well converged numerical results.  

\subsection{Predictions of magic angular momenta}

Foe $N=6$, Figs.\ 1-4 display (in four installments) the total interaction 
energy from EXD as a function of the total angular momentum $L$ in the range 
$ 19 \leq L \leq 140 $. (The total kinetic energy, being a constant, can be
disregarded.) One can immediately observe the appearance of
downward cusps, implying states of enhanced stability, at certain ``magic
angular momenta''. 

For the CF theory, the magic angular momenta can be determined by Eq.\ 
(\ref{ltol*}), if one knows the non-interacting $L^*$'s; 
the CF magic $L$'s in any interval $1/(2m-1) \geq \nu \geq 1/(2m+1)$ 
$[15(2m-1) \leq L \leq 15(2m+1)]$, $m=1,2,3,4,...$,  can be found by adding 
$2mL_0=30m$ units of angular momentum to each of the $L^*$'s. 
To obtain the non-interacting $L^*$'s, one needs first to 
construct\cite{sek,jai1} the compact Slater determinants. The compact 
determinants and the corresponding non-interacting $L^*$'s are listed in 
Table \ref{cfcom}. 

\begin{table}[b]
\caption{\label{cfcom} Compact non-interacting Slater determinants 
and associated angular momenta $L^*$ for $N=6$ electrons according to
the CF presciption. Both $L^*=-3$ and $L^*=3$ are associated with two compact
states each, the one with lowest energy being the preferred one.}
\begin{ruledtabular}
\begin{tabular}{cr}
 Compact state   &   $L^*$   \\ \hline
$[$1,1,1,1,1,1$]$    &   $-$15   \\
$[$2,1,1,1,1$]$      &   $-$9    \\
$[$2,2,1,1$]$        &   $-$5    \\
$[$3,1,1,1$]$        &   $-$3    \\
$[$2,2,2$]$          &   $-$3    \\
$[$3,2,1$]$          &   0       \\
$[$4,1,1$]$          &   3       \\
$[$3,3$]$            &   3       \\
$[$4,2$]$            &   5       \\
$[$5,1$]$           &   9       \\
$[$6$]$              &   15      \\
\end{tabular}
\end{ruledtabular}
\end{table}

\begin{figure}[t]
\centering\includegraphics[width=8.5cm]{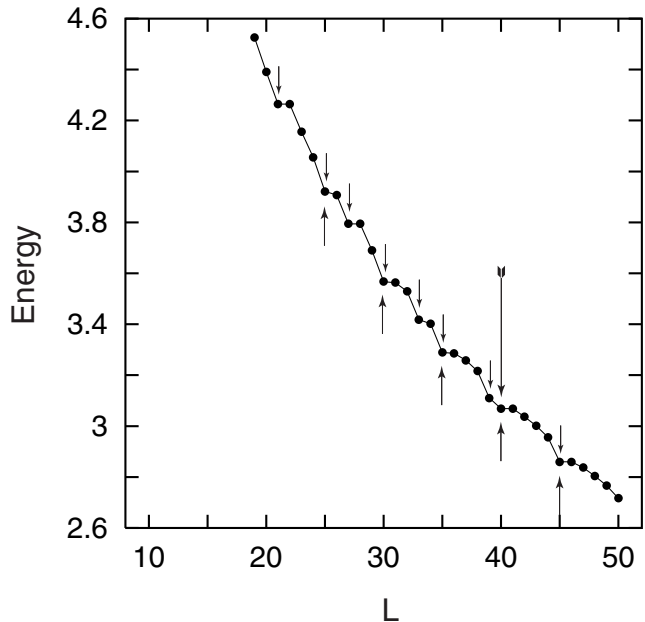}\\
~~~~\\
\caption{
Total interaction energy from exact-diagonalization calculations as a function
of the total angular momentum $(10 \leq L \leq 50)$ for $N=6$ electrons
in high magnetic field. The upwards pointing arrows indicate the magic
angular momenta corresponding to the classically most stable (1,5) polygonal 
ring arrangement of the Wigner molecule. The short downwards pointing arrows 
indicate successful predictions of the composite-fermion model. The
long downward arrow indicates a magic angular momentum not predicted by the
CF model. Energies in units $e^2/\kappa l_B$, where $\kappa$ is the dielectric
constant.
}
\label{ex1}
\end{figure}

There are nine different values of $L^*$'s, and thus the CF theory for $N=6$ 
predicts that there are always nine magic numbers in any interval 
$15(2m-1) \leq L \leq 15(2m+1)$ between two consecutive JL angular momenta 
$15(2m-1)$ and $15(2m+1)$, $m=1,2,3,...$ (henceforth
we will denote this interval as ${\cal I}_m$). For example, using 
Table \ref{cfcom} and Eq.\ (\ref{ltol*}), the CF magic numbers in the 
interval $15 \leq L \leq 45$ ($m=1$) are found to be the following 
nine,\cite{note2}
\begin{equation}
15,\;21,\;25,\;27,\;30,\;33,\;35,\;39,\;45.
\label{cfmam1}
\end{equation}
On the other hand, in the interval $105 \leq L \leq 135$ ($m=4$), the CF 
theory predicts the following set of nine magic numbers,
\begin{equation}
105,\;111,\;115,\;117,\;120,\;123,\;125,\;129,\;135.
\label{cfmam4}
\end{equation}

\begin{figure}[t]
\centering\includegraphics[width=8.5cm]{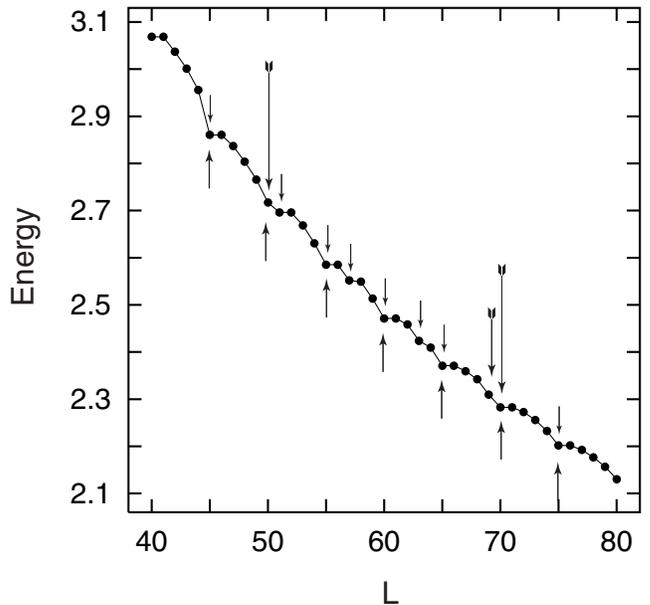}\\
~~~~\\
\caption{
Total interaction energy from exact-diagonalization calculations as a function
of the total angular momentum $(40 \leq L \leq 80)$ for $N=6$ electrons
in high magnetic field. The upwards pointing arrows indicate the magic
angular momenta corresponding to the classically most stable (1,5) polygonal
ring arrangement of the Wigner molecule. The short downwards pointing arrows
indicate successful predictions of the composite-fermion model. 
The medium-size
downwards pointing arrow indicates a prediction of the CF model that fails
to materialize as a magic angular momentum. The long downward arrows indicate 
magic angular momenta not predicted by the CF model. 
Energies in units of $e^2/\kappa l_B$, where $\kappa$ is the dielectric
constant.
}
\label{ex2}
\end{figure}
\begin{figure}[t]
\centering\includegraphics[width=8.5cm]{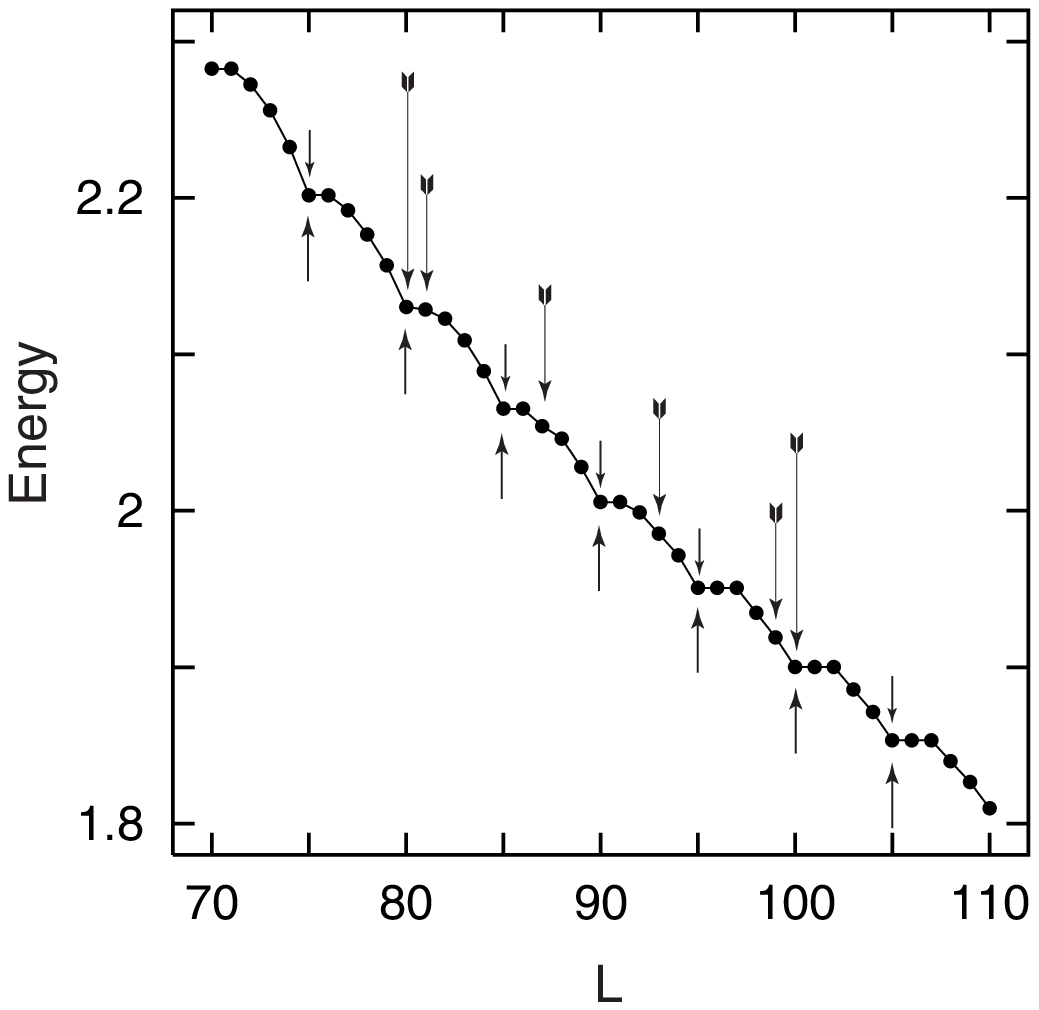}\\
~~~~\\
\caption{
Total interaction energy from exact-diagonalization calculations as a function
of the total angular momentum $(70 \leq L \leq 110)$ for $N=6$ electrons
in high magnetic field. The upwards pointing arrows indicate the magic
angular momenta corresponding to the classically most stable (1,5) polygonal
ring arrangement of the Wigner molecule. The short downwards pointing arrows
indicate successful predictions of the composite-fermion model. 
The medium-size
downwards pointing arrows indicate predictions of the CF model that fail
to materialize as magic angular momenta. The long downward arrows indicate
magic angular momenta not predicted by the CF model.
Energies in units of $e^2/\kappa l_B$, where $\kappa$ is the dielectric
constant.
}
\label{ex3}
\end{figure}
\begin{figure}[t]
\centering\includegraphics[width=8.5cm]{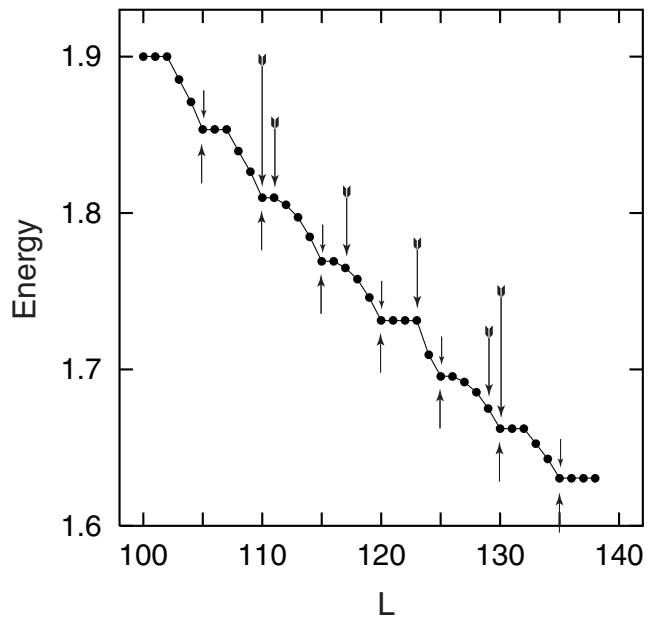}\\
~~~~\\
\caption{
Total interaction energy from exact-diagonalization calculations as a function
of the total angular momentum $(100 \leq L \leq 140)$ for $N=6$ electrons
in high magnetic field. The upwards pointing arrows indicate the magic
angular momenta corresponding to the classically most stable (1,5) polygonal
ring arrangement of the Wigner molecule. The short downwards pointing arrows
indicate successful predictions of the composite-fermion model. 
The medium-size
downwards pointing arrows indicate predictions of the CF model that fail
to materialize as magic angular momenta. The long downward arrows indicate
magic angular momenta not predicted by the CF model.
Energies in units of $e^2/\kappa l_B$, where $\kappa$ is the dielectric
constant.
}
\label{ex4}
\end{figure}

An inspection of the total-energy-vs.-L plots in Figs. 1-4 reveals that the CF 
prediction badly misses the actual magic angular momenta specified by the EXD 
calculations as those associated with the downward cusps. Indeed it is
immediately apparent that the number of downward cusps in any interval
${\cal I}_m$ is always different from 9. Indeed, there are 10 cusps in 
${\cal I}_1$ (including that at $L=15$, not shown in Fig.\ \ref{ex1}),
10 in ${\cal I}_2$ (see Fig.\ \ref{ex2}), 7 in ${\cal I}_3$ (see Fig.\ 
\ref{ex3}), and 7 in ${\cal I}_4$ (see Fig.\ \ref{ex4}). 
In detail, the CF theory fails in the
following two aspects: (I) There are exact magic numbers that are consistently
missing from the CF prediction in every interval; with the exception of the 
lowest $L=20$, these {\it exact\/} magic numbers (marked by a long downward 
arrow in the figures) are given by $L=10(3m-1)$ and $L=10(3m+1)$, 
$m=1,2,3,4,...$; 
(II) There are CF magic numbers that do not correspond to downward cusps in
the EXD calculations (marked by medium-size downward arrows in the figures). 
This happens because cusps associated with $L$'s whose difference from $L_0$ 
is divisible by 6 (but not simultaneously by 5) progressively weaken and 
completely disappear in the intervals ${\cal I}_m$ with $m \geq 3$; only cusps 
with the difference $L-L_0$ divisible by 5 survive.  On the other hand, the CF
model predicts the appearance of four magic numbers with $L-L_0$ divisible 
solely by 6 in every interval ${\cal I}_m$, at $L=30m\mp9$ and $30m\mp3$, 
$m=1,2,3,...$ The overall extent of the inadequacy of the CF model can be 
appreciated better by the fact that there are six false predictions (long and 
medium-size downward arrows) in every interval ${\cal I}_m$ with $m \geq 3$, 
compared to only five correct ones (small downward arrows, see Fig.\ 
\ref{ex3} and Fig.\ \ref{ex4}).

\begin{figure}[t]
\centering\includegraphics[width=7.5cm]{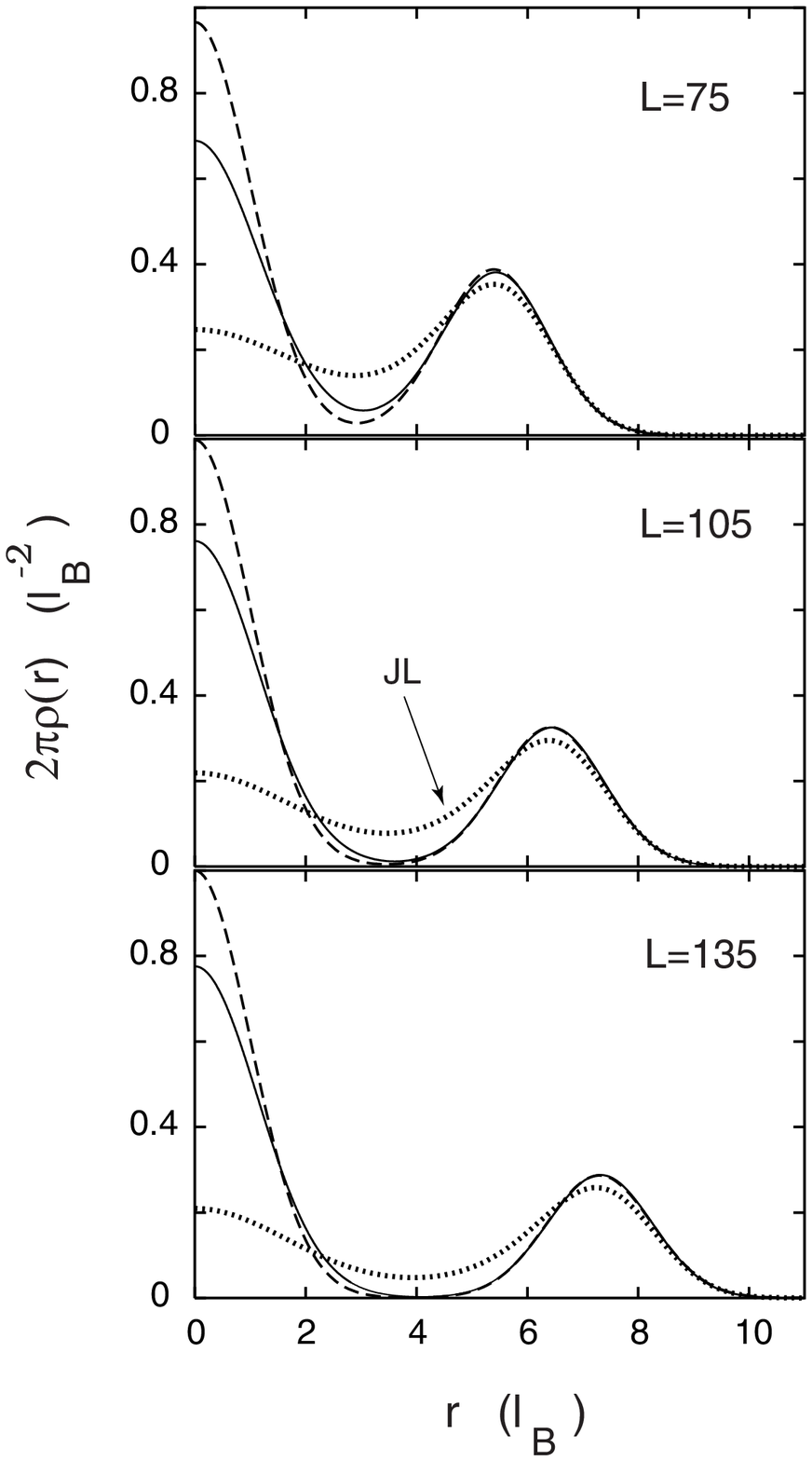}\\
~~~~\\
\caption{
Radial electron densities for $N=6$ electrons in high magnetic field.
Solid line: densities from exact diagonalization. Dashed line: densities from 
REM wave functions. Dotted line: densities from Jastrow-Laughlin wave 
functions.
}
\label{eds}
\end{figure}

In contrast to the CF model, the magic angular momenta in the REM theory
are associated with the polygonal ring configurations of $N$ classical 
point charges. This is due to the fact that the enhanced stability of
the downward cusps results from the coherent collective rotation of the
regular-polygon REM structures. Due to symmetry requirements, such collective 
rotation can take place only at magic-angular-momenta values. The in-between
angular momenta require the excitation of additional degrees of freedom (like
the center of mass and/or vibrational modes), which raises the total energy
with respect to the values associated with the magic angular momenta.

For $N=6$, the lowest in energy ring configuration is the
(1,5), while there exists a (0,6) isomer\cite{bed,bol} with higher energy. 
As a result, our EXD calculations (as well as earlier ones\cite{sek,mak2,rua} 
for lower angular momenta $L \leq 70$) have found that there exist two 
sequences of magic angular momenta, a {\it primary\/} one $(S_p)$ with 
$L=15+5m$ [see Eq.\ (\ref{lwm0n})], associated with the most stable (1,5) 
classical molecular configuration, and a {\it secondary\/} one $(S_s)$ with 
$L=15+6m$ [see Eq.\ (\ref{lwm1nm1})], associated  with the metastable $(0,6)$ 
ring arrangement. Furthermore, our calculations (see also Refs.\  
\onlinecite{rua,mak2}) show that the secondary sequence $S_s$ contributes 
only in a narrow range of the lowest angular momenta; in the region of
higher angular momenta, the primary sequence $S_p$ is the only one that
survives and the magic numbers exhibit a period of five units of angular
momentum. It is interesting to note that the initial competition between the
primary and secondary sequences, and the subsequent prevalence of the
primary one, has been seen in other sizes as well,\cite{rua} i.e., $N=5,7,8$.
Furthermore, this competition is reflected in the field-induced molecular phase
transitions associated with broken symmetry UHF solutions in a parabolic QD. 
Indeed, Ref.\ \onlinecite{sba} demonstrated recently that, as a function of 
increasing $B$, the UHF solutions for $N=6$ first depict the transformation of
the maximum-density-droplet\cite{mac2} into the (0,6) molecular configuration; 
then (at higher $B$) the (1,5) configuration replaces the (0,6) structure
as the one having the lower HF energy.\cite{note3} 

The extensive comparisons in this subsection lead inevitably to the conclusion
that the CF model cannot explain the systematic trends exhibited by the magic 
angular momenta in 2D QD's in high magnetic fields. These trends, however, were
shown to be a natural consequence of the formation of REM's and their 
metastable isomers. 

\subsection{Radial electron densities}

We turn now our attention to a comparison of the radial electron densities
(ED's). Fig.\ \ref{eds} displays the corresponding ED's from EXD, REM, and
CF/JL wave functions at three representative total angular momenta, i.e., 
$L=75$ $(\nu=1/5)$, $105$ $(1/7)$, and $135$ $(1/9)$.

An inspection of Fig.\ \ref{eds} immediately reveals that (I) The EXD radial 
ED's (solid lines) exhibit a prominent oscillation corresponding to the $(1,5)$
molecular structure (averaged over the azimuthal angles). Indeed the integral
of the exact ED's from the origin to the minimum point between the two humps
is practically equal to unity; (II) There is very good agreement between
the REM (dashed lines) and exact ED's; this agreement improves with higher 
angular momentum; (III) The JL ED's (dotted lines ) miss the oscillation of 
the exact ED in all three cases in a substantial way.

The inability of the radial ED's calculated with the JL functions to capture 
the oscillations exhibited by the exact ones was 
also seen recently for the $\nu=1/3$ case and for all electron numbers 
$N=6,7,8,9,10,11,12$ in Ref.\ \onlinecite{tsi} (see in particular Fig.\ 1
therein). We further note that the oscillations of the exact ED's in that
figure correspond fully to the classical molecular ring arrangements 
listed in Ref.\ \onlinecite{bed}, e.g., to (1,7) for $N=8$ and to (3,9) for
$N=12$, in agreement with our rotating-electron-molecule interpretation.

\begin{figure}[t]
\centering\includegraphics[width=7.5cm]{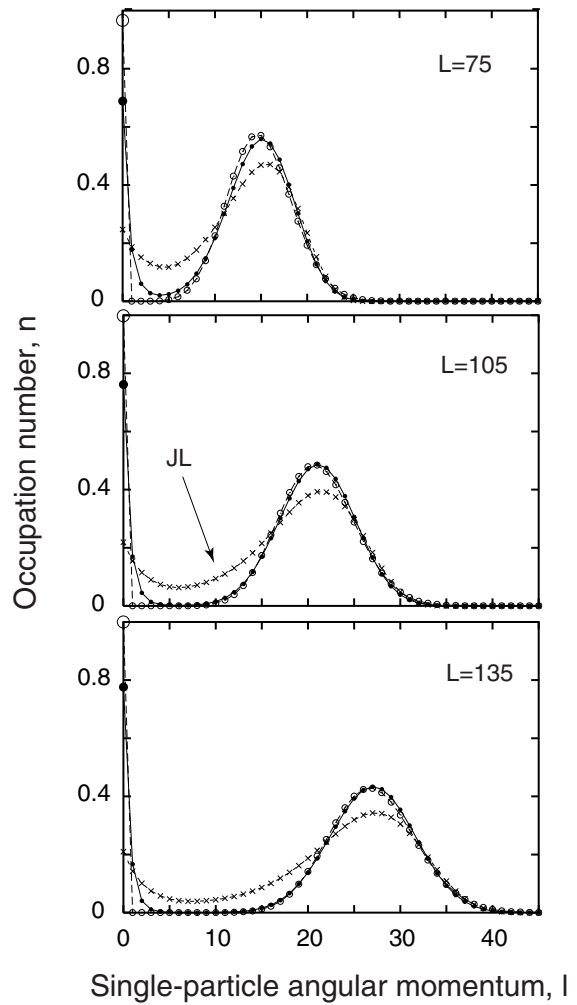}\\
~~~~\\
\caption{
Distribution of occupation numbers as a function of single-particle angular 
momentum $l$ for $N=6$ electrons in high magnetic field.
Solid circles: occupation numbers from exact diagonalization. Open circles: 
occupation numbers from REM wave functions. Crosses: occupation numbers from 
Jastrow-Laughlin wave functions.
}
\label{occs}
\end{figure}

\subsection{Distribution of occupation numbers}

In this subsection, we address the behavior of the occupation-number
distribution $n(l)=\langle \Phi | a^\dagger_l a_l | \Phi \rangle$ as a 
function of the single-particle angular momentum $l$,
where the creation and annihilation operators refer to the single-electron
states $\psi_{0,l}(z)$ in the LLL.  For $N=6$, Fig.\ 
\ref{occs} displays the $n(l)$'s from all three families of wave functions, 
i.e., EXD (solid circles), REM (open circles), and JL (crosses), and for 
the three representative angular momenta $L=75$ $(\nu=1/5)$, $105$ $(1/7)$, 
and $135$ $(1/9)$.

Again, an inspection of Fig.\ \ref{occs} immediately reveals that (I) The 
EXD occupation numbers exhibit a prominent oscillation corresponding to the 
$(1,5)$ molecular structure. Indeed the sum of the exact $n(l)$'s from $l=0$
to the minimum point between the two humps is practically equal to unity; 
(II) There is very good agreement between the REM and exact occupation 
numbers; this agreement improves with higher angular momentum; (III) For
all three cases, the JL occupation numbers exhibit a systematically different 
trend and they are not able to capture the oscillatory behavior of the EXD 
occupation numbers. 

We further note that a substantial discrepancy between JL and EXD 
occupation numbers was also noted in Ref.\ \onlinecite{kash} for the
case of $N=7$ electrons and $\nu=1/3$ $(L=63)$. 

The systematic deviations between the JL and EXD ED's and occupation numbers
inevitably points to the conclusion that these two families of wave functions 
represent very different many-body physical problems. Indeed, the JL 
functions have been found\cite{hk} to be exact solutions for a special class 
of {\it short-range\/} two-body forces, while the EXD functions faithfully 
reflect the {\it long-rang\/}e character of the Coulombic interelectron 
repulsion. On the other hand, as discussed in Ref.\ \onlinecite{yl5}, the REM 
wave functions, derived through a traditional many-body approach,
are able to capture the correlations arising from the long-range 
character of the Coulomb force; the oscillatory behavior of the EXD and REM 
ED's and occupation numbers (associated with formation of Wigner molecules) 
constitutes a prominent and unmistaken signature of such Coulombic 
correlations.

\begin{table}[t]
\caption{\label{over3} 
Case of $N=3$ electrons in high magnetic fileds. Overlaps, 
$\langle \Phi_L|\Phi^{\text{EX}}_L \rangle / 
(\langle \Phi_L|\Phi_L \rangle 
\langle \Phi^{\text{EX}}_L|\Phi^{\text{EX}}_L \rangle)^{1/2}$,
of REM's ($\Phi$'s) and JL functions ($\Phi$'s) with the corresponding exact
eigenstates ($\Phi^{\text{EX}}$'s) for various values of the angular momenta 
$L$ ($\nu$ are the corresponding fractional filling factors).
Recall that the angular momenta for the JL functions are 
$L^{\text{JL}}=N(N-1)(2m+1)/2$, with $m=0,1,2,3,...$ 
The JL overlaps are from Ref.\ \onlinecite{lau1}. }
\begin{ruledtabular}
\begin{tabular}{rcr}
 $L(\nu)$   & JL      & REM     \\ \hline
  9(1/3)    & 0.99946 & 0.98347 \\
 15(1/5)    & 0.99468 & 0.99473 \\
 21(1/7)    & 0.99476 & 0.99674 \\
 27(1/9)    & 0.99573 & 0.99758 \\
 33(1/11)   & 0.99652 & 0.99807 \\
 39(1/13)   & 0.99708 & 0.99839 \\ 
\end{tabular}
\end{ruledtabular}
\end{table}

\subsection{Comparison of overlaps and total energies}

\begin{table}[b]
\caption{\label{over6} Overlaps of JL and REM wave functions with the exact
ones for $N=6$ electrons and various angular momenta L ($\nu$ are the 
corresponding fractional filling factors).}
\begin{ruledtabular}
\begin{tabular}{rcr}
L($\nu$)  & JL          &   REM   \\
\hline
75(1/5)   & 0.837       &  0.817  \\
105(1/7)  & 0.710       &  0.850  \\
135(1/9)  & 0.665       &  0.860  \\
\end{tabular}
\end{ruledtabular}
\end{table}  
We turn now our attention to the overlaps of the REM and JL wave functions 
with those obtained through exact diagonalization. 
We start by listing in Table \ref{over3} the overlaps for the 
simpler case of $N=3$ electrons in high magnetic fields. One sees immediately
that these overlaps are all very close to unity $( \geq 0.99)$ for both
the REM and JL cases and for even rather high angular momenta [e.g., 
$L=39$ $(\nu=1/13)$].

Ever since they were calculated by Laughlin in his original 
paper,\cite{lau1} the JL overlaps for $N=3$ electrons have exercised a great
influence in the literature of the fractional quantum Hall effect (FQHE).
Indeed, in a rather sweeping generalization to any $N$ and $L$
(note that Ref.\ \onlinecite{tsi} has indeed found that the JL overlaps 
for $\nu=1/3$ remain very close to unity for all cases with $5 \leq N \leq 
12$), the close-to-unity values of the JL overlaps have been presumed to 
provide ``proof'' that the CF/JL functions approximate very well the 
corresponding exact many-body wave functions; as we have already shown 
earlier, this presumption is highly questionable.

We have calculated the overlaps for $N=6$ electrons and for the three
representative higher angular-momentum values $L=75$ $(\nu=1/5)$, $105$ 
$(1/7)$, and $135$ $(1/9)$; the results are listed in Table \ref{over6} for 
both the REM and JL wave functions. A most remarkable feature of the results in
Table \ref{over6} is that the extraordinary, higher than 0.99 values (familiar
from Laughlin' s paper\cite{lau1}) are totally absent. Instead, the JL overlaps
rapidly deteriorate for higher $L$'s (lower $\nu$'s), and for $\nu=1/9$
they have attained values below 0.67. In contrast, the REM overlaps remain 
above 0.80 and slowly approach unity as $L$ increases.

From our results for $\nu \leq 1/5$ and the results of Ref.\ \onlinecite{tsi}
for $\nu=1/3$, it is apparent that the overlaps alone are not 
a reliable index for assessing the agreement or disagreement between trial
and exact wave functions. For example, for $N=6$ and $L=75$ $(\nu=1/5)$, Table
\ref{over6} shows that the JL and REM overlaps are close to each other (0.837 
vs. 0.817). However, as the earlier analyses based on the electron densities 
and occupation numbers show, the JL wave function is not a good approximation 
to the exact one; in contrast, the REM wave function offers a much better 
description.

\begin{table}[t]
\caption{\label{toten} Total interaction energies of JL, REM, 
and exact-diagonalization wave functions for $N=6$ 
electrons and various angular momenta L ($\nu$ are 
the corresponding fractional filling factors). The percentages within 
parentheses indicate relative errors. Recall that the angular momenta for the 
JL functions are $L^{\text{JL}}=N(N-1)(2m+1)/2$, $m = 0,1,2,3,...$ 
Energies in units of $e^2/\kappa l_B$, where $\kappa$ is the dielectric
constant.}
\begin{ruledtabular}
\begin{tabular}{lccr}
$L$($\nu$)  & JL          &   REM   & EXACT \\
\hline
75(1/5)   & 2.2093 (0.32\%)   &  2.2207 (0.85\%)   &  2.2018  \\
85(3/17)  & ~~~~              &  2.0785 (0.65\%)   &  2.0651  \\
95(3/19)  & ~~~~              &  1.9614 (0.55\%)   &  1.9506  \\
105(1/7)  & 1.8618 (0.46\%)   &  1.8622 (0.48\%)   &  1.8533  \\
115(3/23) & ~~~~~~~           &  1.7767 (0.45\%)   &  1.7692  \\
125(3/25) & ~~~~~~~           &  1.7020 (0.38\%)   &  1.6956  \\ 
135(1/9)  & 1.6387 (0.50\%)   &  1.6361 (0.34\%)   &  1.6305  \\
\end{tabular}
\end{ruledtabular}
\end{table}  

In addition to the overlaps, earlier studies (see, e.g., Ref.\ 
\onlinecite{jai2}) have also relied on the total energies for assessing the
agreement, or not, between CF and exact wave functions. We thus list in 
Table \ref{toten} the total energies for $N=6$ and for the three 
representative higher angular-momentum values 
$L=75$ $(\nu=1/5)$, $105$ $(1/7)$, and $135$ $(1/9)$.
It is seen that both the JL and REM total energies exhibit very small
relative errors compared to the corresponding EXD ones in all three instances,
a fact that indicates that, by themselves, the total energies\cite{note22} are
an even less reliable index compared to the overlaps. In particular, note that
for $N=6$ and $L=135$, the JL and exact total energies differ only in the
third decimal point, while at the same time the JL overlap is only 0.665
(see Table \ref{over6})!

\subsection{Exponents of current-voltage power law}

Another quantity of theoretical and experimental interest is the ratio
\begin{equation}
\alpha =\frac{ n(l^{\text{JL}}_{\text{max}}-1)}
{ n(l^{\text{JL}}_{\text{max}})},
\label{alpha}
\end{equation}
of the corresponding occupation numbers at $l^{\text{JL}}_{\text{max}}-1$ and 
$l^{\text{JL}}_{\text{max}}$. The interest in this ratio 
is due to the following two facts:
(I) The value of $\alpha$ for the JL function at different fractional fillings 
has a particular analytic value,\cite{mit,wen,gt} i.e., it is given by 
$\alpha^{\text{JL}}(\nu)=1/\nu=2m+1$, $m=1,2,3,4,...$; (II) $\alpha$ happens 
to enter as the exponent\cite{wen,gt} of the voltage in the current-voltage 
law, $I \propto V^\alpha$, for external electron tunneling into an edge of a 
fractional quantum Hall system. Recent investigations have found that both 
the experimental\cite{chan} and computed\cite{gt} EXD 
value of $\alpha$ at $\nu=1/3$ deviates from the JL prediction
of 3, being in all instances somewhat smaller (i.e., $\sim 2.7$).

Table \ref{alphat} displays the values of $\alpha$ for $N=6$ and for 
the JL, REM, and EXD wave functions at various values of the total angular
momentum $L$. We have checked that our numerical values for 
$\alpha^{\text{JL}}$ (derived by dividing the proper $n^{\text{JL}}$'s; see
Fig.\ \ref{occs}) are equal to $2m+1$ within the numerical accuracy. 
As seen from Table \ref{alphat}, a most striking weakness of the JL functions 
is that the corresponding $\alpha^{\text{JL}}$'s 
diverge as $L \rightarrow \infty$, a 
behavior which contrasts sharply with the EXD values that remain at all
times finite and somewhat smaller than 3. Such a dramatic difference in
behavior should be possible to be checked experimentally. Furthermore, we note 
that the REM values, although somewhat smaller, they are close to the EXD ones
and remain bounded as $L \rightarrow \infty$.

We conclude that this dramatic qualitative and quantitative weakness of the JL 
functions is due to their being exact solutions of a family of short range 
interparticle forces.\cite{hk}  
On the other hand, as we have stressed earlier in this paper and in 
Ref.\ \onlinecite{yl5}, the REM functions are able to capture the essential 
effects of the correlations associated with the long-range Coulomb force; thus,
in agreement with the EXD results, the REM $\alpha$ values remain finite
as $L \rightarrow \infty$.

\begin{table}[t]
\caption{\label{alphat} Values of the ratio $\alpha$ [Eq.\ (\ref{alpha})] for
JL, REM, and exact-diagonalization wave functions for $N=6$ electrons and 
various angular momenta L; $\nu$ (given in parentheis) are the corresponding 
fractional filling factors. Recall that the angular momenta for the JL 
functions are $L^{\text{JL}}=N(N-1)(2m+1)/2$, $m = 0,1,2,3,...$ } 
\begin{ruledtabular}
\begin{tabular}{rccr}
L($\nu$)  & JL       &   REM     & EXACT   \\
\hline
75(1/5)   & 5.000    &  1.964   &  2.877  \\
105(1/7)  & 7.000    &  1.972   &  2.708  \\
135(1/9)  & 9.000    &  1.978   &  2.726  \\
\end{tabular}
\end{ruledtabular}
\end{table}

\section{summary}

Exact diagonalization (EXD) results for the lowest rotational band of a 
circular QD with $N=6$ electrons in strong magnetic fields were 
reported\cite{note21} here for the first time in the range of high angular 
momenta $ 70 \leq L \leq 140 $ (covering the corresponding range of fractional
filling factors $1/5 \geq \nu \geq 1/9$). These EXD results were used in a 
thorough assessment of the ability of the 
composite-fermion\cite{jai3}/Jastrow-Laughlin\cite{lau1} and 
rotating-electron-molecule\cite{yl5} trial wave functions to 
approximate the exact wave functions in the case of 2D QD's. 

A detailed comparison (addressing five properties, i.e., prediction of magic 
angular momenta, radial electron densities, 
occupation number distributions, overlaps 
and total energies, and exponents of current-voltage power law) shows that the
REM many-body wave functions provide a description that is superior to that 
obtained through the CF/JL ones. An important finding is that ``global'' 
quantities (like overlaps and total energies) are not particularly reliable 
indices for comparing exact and trial wave functions; a reliable decision on 
the agreement, or lack of it, between exact and trial wave functions should 
include detailed comparisons of quantities like radial electron densities 
and/or occupation number distributions.

We finally note that the CF/JL wave functions have been most useful 
for the modeling of the bulk fractional quantum Hall effect.
However, the theoretical investigations concerning the bulk system
have unavoidably, due to computational limitations, relied on 
finite-size systems to assess the {\it validity\/} of the CF/JL wave 
functions. Thus it is natural to conjecture that the unexpected finding of this
paper, i.e., that the CF/JL functions exhibit remarkable weaknesses in 
reproducing the exact wave functions of QD's in high $B$, may have 
ramifications for our present understanding of the 
fractional quantum Hall effect itself. Investigations of such probable 
ramifications, and related questions concerning the domain of validity
of the REM and CF/JL wave functions in the bulk, will be addressed in future 
publications. In the present paper, we focused on the case of QD's, which
constitute a theoretically self-contained problem when exact-diagonalization 
calculations become available; in the near future, a wider range of such
calculations will be within reach, due to new generations of powerful
computers.

This research is supported by the U.S. D.O.E. (Grant No. FG05-86ER-45234).
Computations were carried out at the Georgia Tech Center for Computational 
Materials Science and the National Energy Research Scientific Computing Center
(NERSC).


\begin{thebibliography}{99}
\bibitem{lau1}
R.B. Laughlin,
Phys. Rev. Lett. {\bf 50}, 1395 (1983).
\bibitem{lau2}
R.B. Laughlin,
Phys. Rev. B {\bf 27}, 3383 (1983).
\bibitem{gir}
S.M. Girvin and T. Jach,
Phys. Rev. B {\bf 28}, 4506 (1983).
\bibitem{jai3}
J.K. Jain,
Phys. Rev. B {\bf 41}, 7653 (1990).
\bibitem{mac}
S.-R. Eric Yang, A.H. MacDonald, and M.D. Johnson,
Phys. Rev. Lett. {\bf 71}, 3194 (1993). 
\bibitem{jai1}
J.K. Jain and T. Kawamura,
Europhys. Lett. {\bf 29}, 321 (1995).
\bibitem{jai4}
T. Kawamura and J.K. Jain,
J. Phys.: Condens. Matter {\bf 8}, 2095 (1996).
\bibitem{mk}
H.-M. M\"{u}ller and S.E. Koonin, 
Phys. Rev. B {\bf 54}, 14532 (1996).
\bibitem{sek}
T. Seki, Y. Kuramoto, and T, Nishino,
J. Phys. Soc. Jpn. {\bf 65}, 3945 (1996).
\bibitem{mak1}
P.A. Maksym,
Phys. Rev. {\bf 53}, 10871 (1996).
\bibitem{rua}
W.Y. Ruan and H-F. Cheung,
J. Phys.: Condens. Matter {\bf 11}, 435 (1999). 
\bibitem{yl2}
C. Yannouleas and U. Landman,
Phys. Rev. B {\bf 61}, 15895 (2000).
\bibitem{mak2}
P.A. Maksym, H. Imamura, G.P. Mallon, and H. Aoki,
J. Phys.: Condens. Matter {\bf 12}, R299 (2000).
\bibitem{cre2m}
C.E. Creffield, J.H. Jefferson, S. Sarkar, and D.L.J. Tipton,
Phys. Rev. B {\bf 62}, 7249 (2000).
\bibitem{yl5}
C. Yannouleas and U. Landman,
Phys. Rev. B {\bf 66}, 115315 (2002).
\bibitem{ron1m}
M. Rontani, G. Goldoni, F. Manghi, and E. Molinari,
Europhys. Lett. {\bf 58}, 555 (2002).
\bibitem{sba}
B. Szafran, S. Bednarek, and J. Adamowski,
Phys. Rev. B {\bf 67}, 045311 (2003).
\bibitem{haw}
See, e.g., A. L. Jacak, P. Hawrylak, and A. Wojs,
{\it Quantum Dots\/} (Springer, Berlin, 1998), 
in particular Ch. 4.5. 
\bibitem{ash}
R.C. Ashoori,
Nature (London) {\bf 379}, 413 (1996).
\bibitem{tar}
S. Tarucha, D.G. Austing, T. Honda, R.J. van der Hage, and L.P. Kouwenhoven,
Phys. Rev. Lett. {\bf 77}, 3613 (1996).
\bibitem{kou}
L.P. Kouwenhoven, C.M. Marcus, P.L. McEuen, S. Tarucha,
R.M. Westervelt, and N.S. Wingreen,
Proceedings of the NATO Advanced Study Institute on {\it Mesoscopic Electron
Transport\/}, Series E, Vol. 345, edited by L.L. Sohn, L.P. Kouwenhoven,
and G. Sch\"{o}n (Kluwer, Dordrecht, 1997) p. 105.
\bibitem{hk}
F.D.M. Haldane, 
Phys. Rev. Lett. {\bf 51}, 605 (1983);
S.A. Trugman and S. Kivelson,
Phys. Rev. B {\bf 31}, 5280 (1985).
\bibitem{note4} 
We use the well-known formula $\nu=N(N-1)/2L$ (see Ref.\ \onlinecite{lau2}), 
which specifies the corresponding fractional filling factors in the 
thermodynamic limit. We stress, however, that in this paper we focus 
exclusively on finite-size systems; thus, throughout this paper, $\nu$ is used
as a more compact index in place of $L$.
\bibitem{yl1}
C. Yannouleas and U. Landman,
Phys. Rev. Lett. {\bf 82}, 5325 (1999);
{\it ibid.\/} {\bf 85}, 2220(E) (2000).
\bibitem{cre1}
C.E. Creffield, W. H\"{a}usler, J.H. Jefferson, and S.Sarkar,
Phys. Rev. B {\bf 59}, 10719 (1999).
\bibitem{gra1} 
R. Egger, W. H\"{a}usler, C.H. Mak, and H. Grabert,
Phys. Rev. Lett {\bf 82}, 3320 (1999);
{\it ibid.\/} {\bf 83}, 462(E) (1999).
\bibitem{yl3}
C. Yannouleas and U. Landman,
Phys. Rev. Lett. {\bf 85}, 1726 (2000).
\bibitem{gra2}
W. H\"{a}usler, B. Reusch, R. Egger, and H. Grabert,
Physica B {\bf 284}, 1772 (2000).
\bibitem{loz1}
A.V. Filinov, Y.E. Lozovik, and M. Bonitz,
Phys. Status Solidi B {\bf 221}, 231 (2000). 
\bibitem{man1}
S.M. Reimann, M. Koskinen, and M. Manninen,
Phys. Rev. B {\bf 62}, 8108 (2000).
\bibitem{gra3}
B. Reusch, W. H\"{a}usler, and H. Grabert,
Phys. Rev. B {\bf 63}, 113313 (2001).
\bibitem{pee1}
A. Matulis and F.M. Peeters,
Solid State Commun. {\bf 117}, 655 (2001). 
\bibitem{loz2}
A.V. Filinov, M. Bonitz, and Y.E. Lozovik,
Phys. Rev. Lett. {\bf 86}, 3851 (2001).
\bibitem{yl4}
C. Yannouleas and U. Landman,
J. Phys.: Condens. Matter {\bf 14}, L591 (2002).
\bibitem{loz3}
P.A. Sundqvist, S.Y. Volkov, Y.E. Lozovik, and M. Willander,
Phys. Rev. B {\bf 66}, 075335 (2002).
\bibitem{mik1}
S.A. Mikhailov and K. Ziegler,
Eur. Phys. J. B {\bf 28}, 117 (2002).
\bibitem{mik2}
S.A. Mikhailov,
Physica E {\bf 12}, 884 (2002).
\bibitem{mik3}
S.A. Mikhailov,
Phys. Rev. B {\bf 65}, 115312 (2002). 
\bibitem{nie1}
A. Harju, S. Siljamaki, and R.M. Nieminen,
Phys. Rev. B {\bf 65}, 075309 (2002).
\bibitem{yl8}
C. Yannouleas and U. Landman,
arXiv: cond-mat/0302130.
\bibitem{bed}
V.M. Bedanov and F.M. Peeters,
Phys. Rev. B {\bf 49}, 2667 (1994).
\bibitem{bol}
F. Bolton and U. R\"{o}ssler,
Superlatt. Microstruct. {\bf 13}, 139 (1993).
\bibitem{note24}
Under conditions of partial spin polarization (i.e., low magnetic fields),
the molecular configurations may exhibit distortions away from the classical
equilibrium configurations. With increasing $R_W$, however, the classical
molecular configurations are recovered (see Ref.\ \onlinecite{mik3}).
\bibitem{jai2}
J.K. Jain and R.K. Kamilla,
Int. J. Mod. Phys. B {\bf 11}, 2621 (1997).
\bibitem{gr}
E. Goldmann and S.R. Renn,
Phys. Rev. B {\bf 56}, 13296 (1997).
\bibitem{yang}
S.-R. Eric Yang and J.H. Han,
Phys. Rev. B {\bf 57}, R12681 (1998).
\bibitem{tau}
M. Taut,
J. Phys.: Condens. Matter {\bf 12}, 3689 (2000). 
\bibitem{man2}
M. Manninen, S. Viefers, M. Koskinen, and S.M. Reimann,
Phys. Rev. B {\bf 64}, 245322 (2001).
\bibitem{rez}
X. Wan, K. Yang, and E.H. Rezayi,
Phys. Rev. Lett. {\bf 88}, 056802 (2002).
\bibitem{rez2}
X. Wan, E.H. Rezayi, and K. Yang,
ArXiv: cond-mat/0302341.
\bibitem{nie2}
A. Harju, S. Siljamaki, and R.M. Nieminen,
Phys. Rev. Lett. {\bf 88}, 226804 (2002).
\bibitem{rs}
P. Ring and P. Schuck,
{\it The Nuclear Many-body Problem\/} (Springer, New York, 1980)
Ch. 11, and references therein.
\bibitem{yl6}
C. Yannouleas and U. Landman,
Eur. Phys. J. D {\bf 16}, 373 (2001).
\bibitem{yl7}
C. Yannouleas and U. Landman,
Int. J. Quantum Chem. {\bf 90}, 699 (2002).
\bibitem{mz}
K. Maki and X. Zotos,
Phys. Rev. B {\bf 28}, 4349 (1983).
\bibitem{note5}
The definition $z \equiv x + \imath y$ is associated with positive angular
momenta for the single-particle states in the lowest Landau level.
In Ref.\ \onlinecite{yl5}, we used $z \equiv x - \imath y$ and negative
single-particle angular momenta in the lowest Landau level. The final
expressions for the trial wave functions do not depend on these choices.
\bibitem{note23}
Both the finite-size REM molecule and several sophisticated bulk Wigner 
crystal (BWC) approaches at high $B$ (listed at the
end of this footnote) start with a single-determinantal UHF wave function
constructed out of the orbitals in Eq.\ (\ref{gaus}), and both do improve it 
by introducing additional correlations; however, the nature of these 
correlations is quite different between the REM and the BWC approaches. 
Indeed, due the the finite-size of the system, the REM approach 
includes correlations associated with fluctuations in the {\it azimuthal\/} 
angle (see Ref.\ \onlinecite{yl5} and Ref.\ \onlinecite{yl8}); 
these correlations arise from the 
restoration of the circular symmetry and result in states with good total 
angular momenta (in particular {\it magic\/} angular momenta, see section 
IV.A). Naturally, in the BWC approaches, angular-momentum conservation and 
magic angular momenta are not considered; for example, Lam and Girvin include 
correlations from {\it vibrational\/}-type fluctuations of the BWC that are 
more in tune with the expected translational invariance of a bulk system.
As a result, the REM exhibits drastically different properties from the 
properties of an N-electron piece of the bulk Wigner crystal. Rather, the
REM wave functions exhibit properties associated with the incompressible
magic-angular-momenta states in the spectra of QD's, which are finite-size 
{\it precursors\/} to the ``correlated-liquid'' fractional quantum Hall states
of the bulk [see Ref.\ \onlinecite{haw}]. 
For sophisticated BWC approaches at high $B$, see, e.g., P.K. Lam and 
S.M. Girvin, Phys. Rev. B {\bf 30}, 473 (1984); H. Yi and H.A. Fertig,
Phys. Rev. B {\bf 58}, 4019 (1998).
\bibitem{df}
C.G. Darwin, Proc. Cambridge Philos. Soc. {\bf 27}, 86 (1930);
V. Fock, Z. Phys. {\bf 47}, 446 (1928).
\bibitem{note1}
In the case of $N=6$ electrons, we have used 338, 5444, 32134, and 118765 terms
in this decomposition for $L=45\; (\nu=1/3)$, $75 \;(1/5)$, $105\; 
(1/7)$, and $135\;(1/9)$, respectively. These numbers correspond to
all the Slater determinants with $L=15(2m+1)$ and individual angular momenta
$l \leq l^{\text{JL}}_{\text{max}}=5(2m+1)$, including the cases with zero 
coefficients. We remind the reader that $l^{\text{JL}}_{\text{max}}=
(2m+1)(N-1)$ is the maximum individual angular momentum allowed in the JL 
states. The Slater decomposition of the JL states for $N=2,3,4,5,6$, 
but only for $\nu=1/3$, has been reported earlier in 
G.V. Dunne, Int. J. Mod. Phys. B {\bf 7}, 4783 (1993). 
\bibitem{math}
S. Wolfram, {\it Mathematica: A system for doing mathematics
by computer\/} (Addison-Wesley, Reading, 1991).
\bibitem{sto}
M. Stone, H.W. Wyld, and R.L. Schult,
Phys. Rev. B {\bf 45}, 14156 (1992).
\bibitem{tsi}
E.V. Tsiper and V.J. Goldman,
Phys. Rev. B {\bf 64}, 165311 (2001).
\bibitem{kasn}
M. Kasner,
Ann. Phys. (Berlin) {\bf 11}, 175 (2002),
and references therein.
\bibitem{kash}
V.A. Kashurnikov, N.V. Prokof'ev, B.V. Svistunov, and I.S. Tupitsyn,
Phys. Rev. B {\bf 54}, 8644 (1996).
\bibitem{tsi2}
E.V. Tsiper,
J. Math. Phys. {\bf 43}, 1664 (2002). 
\bibitem{arp}
R.B. Lehoucq, D.C. Sorensen, and C. Yang,
{\it ARPACK Users' Guide: Solution of Large-Scale Eigenvalue Problems
with Implicitly Restarted Arnoldi Methods\/}
(SIAM, Philadelphia, 1998).
\bibitem{note2}
Ref.\ \onlinecite{sek} gives the full list of the nine CF magic numbers in the
interval $(1 \geq \nu \geq 1/3)$. Ref.\ \onlinecite{jai1} excludes two of 
them, i.e., the CF magic angular momenta 27 and 33.
\bibitem{mac2}
A.H. MacDonald, S.R.E. Yang, and M.D. Johnson, 
Aust. J. Phys. {\bf 46}, 345 (1993).
\bibitem{note3}
At $B=0$, the interelectron Coulombic repulsion can also induce (as a function
of increasing $R_W$) a similar succession of phase transitions [i,e., normal 
fluid $\rightarrow$ (0,6) molecule $\rightarrow$ (1,5) molecule],
see Fig.\ 2 in Ref.\ \onlinecite{yl1}.
\bibitem{note22}
There is no ``variational dilemma'' from the fact that the CF/JL and REM
functions are two essentially different wave functions with very close
expectation values of the energy. Indeed, the CF/JL wave functions correspond 
to a hamiltonian with short-range two-dody interactions, while the REM 
functions correspond to the actual hamiltonian of the Coulomb problem that 
involves long-range interelectron interactions. Therefore these represent two 
separate variational problems.
\bibitem{mit}
S. Mitra and A.H. MacDonald,
Phys. Rev. B {\bf 48}, 2005 (1993).
\bibitem{wen}
X.G. Wen, Phys. Rev. B {\bf 41}, 12838 (1990); Int. J. Mod.
Phys. B {\bf 6}, 1711 (1992).
\bibitem{gt}
V.J. Goldman and E.V. Tsiper,
Phys. Rev. Lett. {\bf 86}, 5841 (2001). 
\bibitem{chan}
A.M. Chang, M.K. Wu, C.C. Chi, L.N. Pfeiffer, and K.W. West,
Phys. Rev. Lett. {\bf 86}, 143 (2001), and references therein.
\bibitem{note21}
We have specifically considered the limit when the confining potential can be 
neglected compared to the confinement induced by the high magnetic field. 
In high $B$, the effect of the confining potential amounts simply in selecting
a specific magic-angular-momentum state (see section IV.A) as the ground state
of the system (the specific value of the magic $L$ depends on the strength of 
$B$ and the parameters of the confinement). In most studies (see, e.g., 
Ref.\ \onlinecite{jai1}, Ref.\ \onlinecite{sek}, or Ref.\ \onlinecite{mak2}), 
the external confinement has been modeled as a harmonic potential.
Most recently, however, Wan {\it et al.\/} 
(Ref.\ \onlinecite{rez} and Ref.\ \onlinecite{rez2}) have studied few-electron 
QD's taking into consideration a disk-like neutralizing positive background.
Indeed, these authors employ a confining potential arising from a 
positive background charge distributed uniformly on a parallel disk at a 
distance $d$ from the electron layer (the typical $d$ in experiments is $d 
\geq 10 l_B$, see Ref.\ \onlinecite{rez2}). As was the case with the harmonic
external potential, these authors found 
again that their external confinement influences which magic-$L$ state becomes 
the ground state of the system. Most importantly, the ground-state wave 
functions in their exact-diagonalization study exhibit strong oscillations 
in the radial electron density [in an apparent agreement with the classical
ring configurations of Wigner molecules] and in disagreement with the CF/JL 
wave functions. It is interesting to note the coincidence, for all practical 
purposes, of the exact radial electron density for $N=6$ and $L=105$  
calulated by Wan {\it et al.\/} with that calculated by us [compare figure 
5(d) in Ref.\ \onlinecite{rez2} with the middle panel of figure 5 in this 
paper]. In order to account for the disagreement between the exact and CF/JL 
wave functions, Wan {\it et al.\/} were led to use the concept of ``edge 
reconstruction''. In the case studied by us, however, our 
exact-diagonalization results (and those of 
Tsiper and Goldman, see Ref.\ \onlinecite{tsi}) do not 
include any external confinement, a fact that rules out ``edge 
reconstruction'' as the underlying cause for the disagreement between the 
exact and CF/JL wave functions. As we have pointed out in this paper 
previously (see section IV.E and also Ref.\ \onlinecite{yl5}), this 
disagreement arises from the fact that the CF/JL functions do not capture the 
long-range character of the Coulomb interelectron repulsion. On the contrary 
the REM wave functions are able to capture the long-range Coulombic 
correlations and thus are in better agreement with the wave functions from 
exact diagonalization. 

\end{thebibliography}
\end{document}